\documentclass[aps,pre,twocolumn,showpacs,superscriptaddress]{revtex4}
\usepackage{epsfig}
\usepackage{amsmath}
\usepackage{color}
\begin{document}

\title{Stringent Numerical Test of the Poisson Distribution for Finite Quantum
Integrable Hamiltonians}

\author{A. Rela\~no}
\affiliation{Departamento de F\'{\i}sica At\'omica, Molecular y Nuclear,
         Universidad Complutense de Madrid,
         E-28040 Madrid, Spain}

\author{J. Dukelsky}
\affiliation{Instituto de Estructura de la Materia, CSIC,
             Serrano 123,  28006 Madrid, Spain}

\author{J. M. G. G\'omez}
\affiliation{Departamento de F\'{\i}sica At\'omica, Molecular y Nuclear,
         Universidad Complutense de Madrid,
         E-28040 Madrid, Spain}

\author{J. Retamosa}
\affiliation{Departamento de F\'{\i}sica At\'omica, Molecular y Nuclear,
         Universidad Complutense de Madrid,
         E-28040 Madrid, Spain}

\begin{abstract}

Using a new class of exactly solvable models based on the pairing
interaction, we show that it is possible to construct integrable
Hamiltonians with a Wigner distribution of nearest neighbor level
spacings. However, these Hamiltonians involve many-body interactions
and the addition of a small integrable perturbation very quickly leads
the system to a Poisson distribution. Besides this exceptional cases,
we show that the accumulated distribution of an ensemble of random
integrable two-body pairing hamiltonians is in perfect agreement with
the Poisson limit. These numerical results for quantum integrable
Hamiltonians provide a further empirical confirmation to the work of
the Berry and Tabor in the semiclassical limit.

\end{abstract}


\maketitle

%
%

The concept of quantum chaos still lacks a clear definition.  The main
ideas in this field have been obtained using the semiclassical
approximation for quantum systems having a classical analogue. In a
seminal paper, Bohigas {\it et al.} \cite{Bohigas:84} conjectured that
the fluctuation properties of generic quantum systems, which in the
classical limit are fully chaotic, coincide with those of random
matrix theory (RMT). This conjecture is strongly supported by
experimental data, numerical calculations, and analytical work based
on semiclassical arguments. For a generic quantum integrable system,
Berry and Tabor \cite{Berry:77} showed that under very general
conditions the spectral fluctuations in the semiclassical limit are
well described by the Poisson statistics, i. e.  the successive energy
levels are uncorrelated.

The analysis of spectral fluctuations provides an essential tool in
the study of quantum chaos.  Moreover, as RMT was introduced to
explain the fluctuation properties of many-body systems like the
atomic nucleus, it is usually considered that the level statistics
establishes a link between many-body systems and the semiclassical
picture. Actually, if the semiclassical limit is not valid, the
comparison of the system spectral fluctuations with those predicted by
RMT is the main criterion to decide whether the system dynamic is
regular or chaotic. When spectral fluctuations fall very near the RMT
predictions the quantum system is considered fully chaotic. On the
contrary, if they follow closely enough the Poisson statistic, the
system is considered regular. Therefore, the concepts of chaotic and
regular quantum systems are not well established since they rely on
results that have only been checked in the semiclassical limit.

The concept of integrability in Classical Mechanics is well defined
after the work of Liouville in the 19th century. A classical
Hamiltonian system is integrable if it has a complete set of
independent integrals of motion commuting with respect to the Poisson
brackets. The total number of integrals of motion should be half of
the dimension of the phase space to assure completeness. In Quantum
Mechanics, the concept of integrability is usually derived from an
extension of the Liouville's definition. A quantum system is said to
be integrable if it is possible to define a complete set of hermitian
operators, the integrals of motion, that commute among each
other. However, this definition has some deficiencies and
ambiguities. Various attempts to clarify the characteristics of these
integrals of motions, mainly their functional independency, have not
produced irrefutable answers (see, for instance, \cite{weigert:92,
weigert:95, Feng:95} and references therein). In this work we shall
use the following definition: {\it a quantum system is said to be
integrable if a set of as many commuting hermitian operators as
quantum degrees of freedom can be explicitly given, and the
hamiltonian can be expressed as a function of these operators}
\cite{weigert:95}. This criterion, relying in the algebraic structure
of Quantum Mechanics, does not directly refer to Classical
Mechanics. Since this definition requires that a ``complete'' set of
quantum integrals of motion is explicitly given, it assures the
existence of a common basis of eigenstates labelled by the eigenvalues
of the integrals of motion. Moreover, the system is Exactly Solvable
if the complete set of eigenstates can be found by algebraic
methods. In this sense, the previous definition of quantum
integrability is closely related to exact solvability. We will use
this criterion to test some accepted properties of quantum integrable
systems, like the Berry and Tabor conjecture, without taking into
account the classical limit.

The level statistics of quantum integrable models has been analyzed in
condensed-matter physics as well as in nuclear physics. Poilblanc {\em
et al.}  \cite{Poilblanc:93} studied the spectral fluctuation by
finding the energy spectrum for several one-dimensional finite lattice
models, like the Heisenberg model, the $t-J$ model and the Hubbard
model. Alhassid and Novoselsky \cite{Alhassid:92} studied the quantal
fluctuations displayed by the energy levels in the Interacting Boson
Model of nuclei.  Recently, the realization of a Poisson distribution
has been suggested as a detector of new integrable quantum chiral
Potts models \cite{Maillard}. In each case, the integrable Hamiltonian
is parameter independent, or it depends on a single free parameter
like in the Hubbard model (the on-site repulsion U). The study of the
spectral fluctuations has been carried out diagonalizing a definite
Hamiltonian in the largest possible Hilbert apace. In all cases of
integrable quantum models, it was verified that the histogram of the
near-neighbors level spacing could be well fitted by a Poisson
distribution. These results provide a numerical support for the Berry
and Tabor semiclassical demonstration \cite{Poilblanc:93, Alhassid:92,
Maillard}, but the quality of the statistics is poor due to the small
number of energy levels entering in the histogram. This is in contrast
with the numerical testing of the Bohigas \cite{Bohigas:84}
conjecture, assigning a Wigner distribution to any non-integrable
quantum hamiltonian, in which large ensembles of two-body random
Hamiltonians were considered (for a review see \cite{Kota}).  More
recently, Benet {\em et al.} \cite{Benet:03} have studied an ensemble
of integrable bosonic hamiltonians whose members display GOE or
GUE-like spectra with probability one. This anomalous behavior can be
explained in terms of semiclassical mechanics. Despite of the fact
that the systems under consideration were integrable (in the
semiclassical sense), the periodic orbits that fulfill the
Einstein-Brillouin-Keller (EBK) quatization condition explore huge
regions of the phase space, i. e. they mimic a typical chaotic motion;
thus, it is reasonable to find random matrix spectral
fluctuations. Therefore, the whole ensemble can be considered an
anomalous exception of Berry and Tabor conjecture.

Trying to get more insight on these ideas, we study the level
statistic of a new class of quantum integrable models: the
Richardson-Gaudin models; they are based on the pairing interaction
\cite{Dukelsky:01} and have a large number of free parameters which
can be selected randomly. In particular, we will study the rational
model for which the quantum invariants have the form

\begin{eqnarray}
R_{i} &=&K_{i}^{0}+ 2g\sum_{j\left( \neq i\right) }\frac{1}{\eta_i
- \eta_j} \left\{\frac{1}{2}\left( K_{i}^{+}K_{j}^{-}+K_{i}^
{-}K_{j}^{+}\right) \right.  \nonumber \\
&&\left. + K_{i}^{0}K_{j}^{0}\right\}, \label{Oper}
\end{eqnarray}
where $i$ labels the $\Omega$ levels of a single particle basis,
$\eta_i$ are $\Omega$ free real parameters and $g$ is the paring
strength. The operators $ K^+, K^-$ and $K^0$ are the $SU(2)$
generators of the pair algebra in level $i$

\begin{equation}
K_{i}^0=\frac{1}{2} ( a_{i}^{\dagger
}a_{i}+a_{\overline{i}}^{\dagger }a_{\overline{i}}-1)\quad ,\quad
K_{i}^{+}= \frac{1}{2} a_{i}^{\dagger }a_{\overline{i}}^{\dagger
}=\left( K_{i}^-\right) ^{\dagger }. \label{ope}
\end{equation}

The operator $K^+_i$ creates a pair of particles in time reversal
states in the double degenerated level $i$. The three generator
close the commutation algebra $SU(2)$

\begin{equation}
\left[ K_{l}^{0},K_{l^{\prime }}^{+}\right] =\delta _{ll^{\prime
}}K_{l}^{+}~,\quad \left[ K_{l}^{+},K_{l^{\prime }}^{-}\right] =
2\delta _{ll^{\prime }}K_{l}^{0}.  \label{SU}
\end{equation}

Based on this commutation algebra, it is straightforward to check that
the operators (\ref{Oper}) commute with one another for arbitrary
values of $g$ and the set of $\Omega$ parameters $\eta$.  The rational
model, as well as other models, is fully integrable and exactly
solvable (for the exact eigenstates of these models see
\cite{Dukelsky:01}).

Once the free parameters inside the $R_i$ operators are fixed, it
is possible to find their complete set of common eigenvalues
$r_i^\alpha$ and eigenvectors $|\psi_\alpha>$. Any function of the
$R_i$ operators defines a valid integrable Hamiltonian. In
particular, linear combinations of the $R_i$ operators produce
quite general pairing hamiltonians

\begin{eqnarray}
H = \sum_i\varepsilon_i R_i =
\sum_i \varepsilon_i K_{i}^{0} + \nonumber \\
g \sum_{j \left( \neq i\right) }\frac{\varepsilon_1 - \varepsilon_j}{\eta_i - \eta_j}
\left\{\frac{1}{2}\left( K_{i}^{+}K_{j}^{-}+K_{i}^{-}K_{j}^{+}\right) \right.
\left. + K_{i}^{0}K_{j}^{0}\right\}. \label{Hoper}
\end{eqnarray}

Contrary to most of the integrable Hamiltonians cited in the
introduction, these new class of integrable Hamiltonians depend on
$(2\Omega+1)$ free, real and independent parameters; combinations of
higher rank quantum invariants would give larger sets of free
parameters.  Within this very large parameter space it is worth to
explore the possibility of finding chaotic spectra, contradicting the
Berry and Tabor hypothesis. Therefore the usual distinction between
chaos and regularity by means spectral statistics would become
doubtful.

To shed some light on this question we have tried to fit several
chaotic spectra using the class of Hamiltonians (\ref{Hoper}) for
different values of $\Omega$ and different particle pair numbers $N$.
These chaotic spectra were obtained by diagonalizing a random matrix
with the appropriate dimension.  The dimension of the Hilbert space
for a system with $\Omega$ levels and $N$ fermion pairs is given by
$D=\displaystyle \binom{\Omega}{N}$. Thus, is usually much larger than
the number of Hamiltonian parameters $2\Omega+1$; therefore it is
impossible to obtain an exact replica of the actual random matrix
spectrum, and we can only get an approximation to this spectrum.  In
all the cases the best parameter set leads to a Hamiltonian whose
spectrum shows Poisson level fluctuations, as predicted by Berry and
Tabor.

Nevertheless, as we have commented above, it is still possible to
consider more general Hamiltonians using the $R_i$ operators. These
Hamiltonians involve many body forces represented by combinations of
higher rank $R$ operators. Knowing the dimension $D$ of the Hilbert
space for a system with $\Omega$ levels and $N$ fermion pairs, we
propose the following class of integrable Hamiltonians with many body
forces.

\begin{equation}
H = \sum_{i_1 > i_2 > \cdots > i_N} \varepsilon_{i_1,i_2,\cdots
,i_N} R_{i_1} R_{i_2} \cdots R_{i_N}. \label{HR}
\end{equation}

\noindent

If we fix the $\eta_i$ parameters and the value of $g$ inside the
$R_i$ operators entering in (\ref{HR}), the number of free parameters
$\varepsilon_{i_1,i_2,\cdots,i_N}$ in the hamiltonian is precisely
equal to the dimension of the Hilbert space $D$.  Therefore, the $d$
eigenvalues of a non-integrable pairing Hamiltonian with a typical
chaotic spectrum can be exactly fitted with the Hamiltonian (\ref{HR})
by solving a linear set of equations for the $D$ unknowns
$\varepsilon_{i_1,i_2,\cdots,i_N}$.

We have considered several cases with $(\Omega,N) = (10,6)$
$(11,6)$,$(12,6)$ and $(13,6)$; the corresponding dimensions of the
Hamiltonian matrices are $D = 210$, $462$, $924$, $1716$
respectively. In each case we have been able to fit a chaotic spectra
with the appropriate dimensionality obtaining an exception to the
Berry and Tabor result. In other words, we have been able to obtain
spectral fluctuations of Wigner-Dyson type at all the energy scales
for quantum integrable hamiltonians. Actually, exceptions to this rule
were already known: for example, Crehan \cite{Crehan:95} proved that
any spectral sequence obeying a certain growth restriction is the
quantum spectrum of an equivalence class of classically integrable
nonlinear oscillators. Our result, however, is quite more enlightening
because it provides an example of a full and realistic quantum system
where the semiclassical result of Berry and Tabor does not apply.

Our findings rise the question of how stable are these solutions
against small perturbations of the Hamiltonian parameters within the
parameter space and whether finite size effects may affect these
properties. To study these two points the Hamiltonians obtained in the
previous fit are perturbed as follows. Each parameter
$\varepsilon_{i_1,i_2,\cdots,i_N}$ in (\ref{HR}) is replaced by

\begin{equation}
\varepsilon \longrightarrow \varepsilon^{'} = \varepsilon
(1+\phi\lambda), \label{pert}
\end{equation}

\noindent
where $\lambda$ stands for the perturbation strength and $\phi$ is a
phase chosen at random. Notice that the new Hamiltonian is also a
combination of the $R_i$ operators and therefore it is integrable.

Before we proceed to analyze the spectral fluctuations of the
perturbed Hamiltonians, their spectra must be unfolded.  For any
quantum system the level density $\rho(E)$ can be separated into a
smooth part $\overline{\rho(E)}$ and a fluctuating part
$\widetilde{\rho(E)}$. The former gives the main trend of the level
density and the later characterizes the spectral
fluctuations. Similarly, the cumulative level density, that gives the
true number of levels up to energy E,
\begin{equation}
N(E) = \int_{-\infty}^E dE' \rho(E'),
\end{equation}
can be separated into a smooth part $\overline{\rho(E)}$ and a
fluctuating $\widetilde{\rho(E)}$ part, i.e.,
\begin{equation}
N(E) =  \overline{N(E)}+\widetilde{N(E)}.
\end{equation}
Actually, level fluctuation amplitudes are modulated by the value of
the mean level density $\overline{\rho(E)}$; therefore, to compare the
fluctuations of different systems, or even the fluctuations of
different parts of the same spectrum, the level density smooth
behavior must be removed. This is done by means of a transformation,
called unfolding \cite{Haake:01}, which consists in mapping the energy
levels $E_i$ onto new dimensionless $\zeta_i = \overline{N(E)}$.
Then, the nearest neighbor spacing sequence is defined by
\begin{equation}
s_i=\zeta_{i+1} - \zeta_i,\,\,\,i=1,\cdots,N-1.
\end{equation}
For the unfolded levels $\bar{\rho}(\epsilon)=1$ and
$\left<s\right>=1$.  In general this is a difficult task for systems
where an analytical expression for the mean level density is not
known.  This is the case of the hamiltonian ensembles introduced in
the present Letter. Therefore, in order to obtain a good approximation
to $\overline{N(E)}$ we have performed a least square fit to Chebyshev
polynomials.

The spectral properties of the perturbed Hamiltonians can be analyzed
by means of different statistics. The most simple is the nearest
neighbor spacing distribution $P(s)$.  It allows to quantify the
chaoticity of the system in terms of a single parameter $\omega$ by
fitting the $P(s)$ histogram to the Brody distribution $P(s,\omega)$
\cite{Brody:81} using a least-squares method. The resulting
distribution interpolates between the Poisson limit $(\omega = 0)$ and
the Wigner limit $(\omega=1)$. It would be also possible to analyze
the behavior of the eigenstates as a function of the perturbation
strength. This could be done calculating the information entropy $I_H$
or the localization length $l_H$ \cite{Izrailev:89}. However, we can
advance that the Hamiltonian eigenstates will remain unaltered under
the influence of the perturbation (\ref{pert}) which is defined on the
coefficients of the combination of integrals of motion but does not
modify them. Actually, for any integrable system the eigenstates are
completely defined by the quantum integrals of motions, and the
Hamiltonian can be expressed as a function of these
operators. Therefore, all Hamiltonians obtained using different
functions of a given complete set of integrals of motion will have the
same eigenstates.

\begin{figure}[t]
\begin{center}
\leavevmode
\epsfig{file=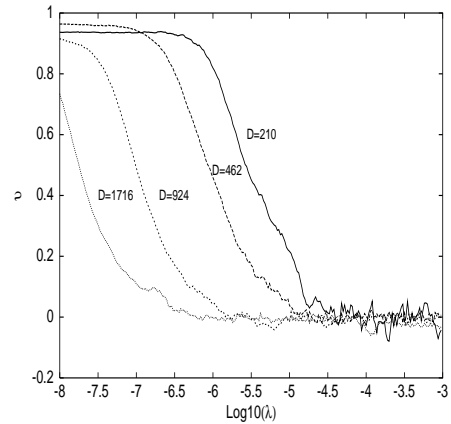,height=6cm,width=6cm,angle=-90}
\caption[]{Variation of the Brody parameter $\omega$ as a function of
the perturbation strength $\lambda$ for $(\Omega,N) = (10,6)$
$(11,6)$,$(12,6)$ and $(13,6)$.}
\label{fig:6cue}
\end{center}
\end{figure}

Fig. \ref{fig:6cue} displays in a semi-logarithmic scale the Brody
parameter $\omega$ as a function of the perturbation strength
$\lambda$ for the four $(\Omega,N)$ examples we have considered
before.  In all the cases a very small perturbation is enough to drive
the system to the Poisson limit. For the smallest system we considered
($D=210$) a perturbation $\lambda \sim 10^{-5}$ is enough to obtain
poissonian spectral fluctuations, while for the largest system
($D=1716$) three order of magnitude less are required. Clearly, the
trend is that larger systems require smaller perturbations.

On the light of these results we conclude that it is necessary to
consider integrable hamiltonians with many-body interactions in order
to obtain a chaotic-like energy spectrum. However, small perturbations
within the integrable space of parameters restore the poissonian like
spectrum. We conjecture that for very large $\Omega$ and N values a
chaotic spectra would correspond to isolated points in the parameter
space and that a infinitesimal perturbation within this space would
immediately drive the system to a Poisson distribution.

In some cases it is possible to introduce a suitable random matrix
model that describes the behavior of spectral fluctuations as the
system evolves through the parameter space.  This has already been
done for the metal-insulator transition or for the order-disorder
transition in Quantum Hall systems \cite{Shukla:00}. However, in the
prersent case, this approach seems to be more complicated since the
``Wigner phase'' corresponds to isolated points in the parameter
space.

In view of the previous results, one would expect that a more
physically sound family of pairing Hamiltonians with two body forces
like (\ref{Hoper}) should give rise to a clean Poisson level
statistics. In order to verify the correctness of this statement we
have studied an ensemble of Hamiltonians (\ref{Hoper}) that we shall
call ``Two Body Pairing Random Ensemble'' ({\bf TBPRE}). We selected
the case of $(\Omega,N) = (13,6)$, corresponding to largest dimension
$D=1716$. For the sake of simplicity, the $\eta_i$ parameters and the
strength $g$ were given fix values and the $\varepsilon_i$
coefficients were selected randomly. The quantum invariants $R_i$
(\ref{Oper}) are independent of the $\varepsilon_i$ parameters and
thus, their corresponding eigenvalues stay the same for the whole
ensemble. In our calculations we have used $200$ ensemble members.

\begin{figure}[h]
\begin{center}
\leavevmode \epsfig{file=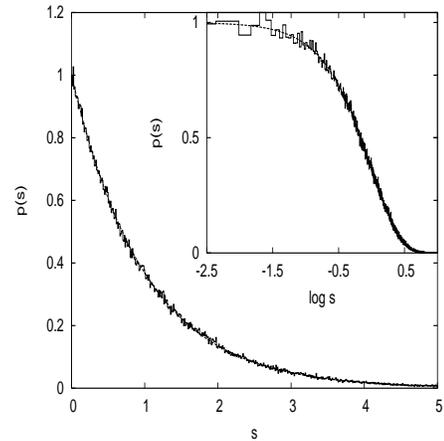,height=6cm,width=6cm,angle=-90}
\caption[]{Nearest neighbor spacing distribution, $P(s)$, for 200
{\bf TBPRE} members. The dashed curve correspond to the Poisson limit.}
\label{fig:p(s)}
\end{center}
\end{figure}

\begin{figure}[ht]
\begin{center}
\leavevmode
\epsfig{file=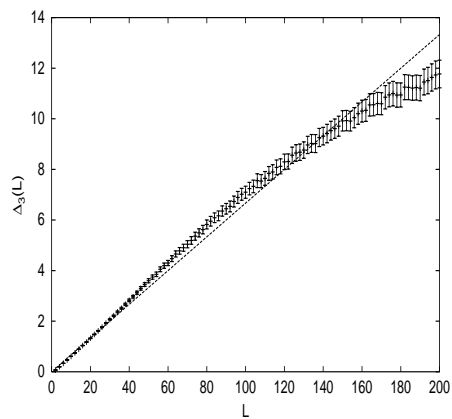,height=6cm,width=6cm,angle=-90}
\caption[]{$\overline{\Delta_3(L)}$ statistic calculated for 200
{\bf TBPRE} members. The dashed line represents the Poisson values.}
\label{fig:delta3}
\end{center}
\end{figure}

The short and long range spectral correlations of this ensemble have
been analyzed by means of the usual level statistics distribution
$P(s)$ and by the rigidity $\Delta_3(L)$ respectively. In the Poisson
limit, characteristic of a regular system, the nearest neighbor
spacing distribution behaves as $P^{Poisson}(s) = \exp{(-s)}$ and
$\Delta^{Poisson}_3(L) = L/15$. Fig. \ref{fig:p(s)} compares the
nearest neighbor distribution $P(s)$, calculated numerically for our
ensemble with the expected Poisson limit. We present the results in
normal as well as in semi-logarithmic scales to enlarge the small
spacing region, which has been shown to present some deviations from
the Poisson limit \cite{casati:85}. It can be seen that the histogram
and the theoretical curve match perfectly. Fig. \ref{fig:delta3} shows
the calculated $\Delta_3(L)$ and compares it to the Poisson limit
$L/15$. The agreement is almost perfect up to $L\simeq 40$; for larger
$L$ values the $\Delta_3$ shows an slight upbending from the Poisson
straight line.  The calculation of the rigidity is quite sensitive to
unfolding procedure. When the mean level density is not accurate
enough known, the unfolding method will introduce accumulated errors
that eventually give rise to a spurious increase of the $\Delta_3(L)$
for large $L$ values \cite{Gomez:02}. The results shown in these two
figures make it possible to conclude that the TBPRE spectral
fluctuations are very well described by the Poisson
statistic. Actually, Figures 2 and 3 constitute the most precise
numerical verification of the Berry and Tabor theoretical proof due to
the fact that we were able to accumulate statistics by using an
ensemble of random integrable Hamiltonians ({\bf TBPRE}) which, to the
best of our knowledge, it wouldn't be possible for any other
integrable model.

To support in a more precise way this conclusion we consider again the
$p(s)$ statistics. As this statistics is less sensitive to the
unfolding procedure than the $\Delta_3$, small deviations from the
theoretical Poisson prediction can be connected to the actual
characteristics of the system dynamics. To avoid any effects related
to the bin size in the $P(s)$ histogram, we will use the accumulated
nearest neighbor spacing distribution $I(s) = \int_0^s P(s')ds'$,
which in Poisson limit it is given by $I^{Poisson}(s) = 1 -
\exp{(-s)}$.  We define a ``distance'' between the calculated $n(s)$
distribution and the Poisson limit as

\begin{equation}
\delta^2 = \int_0^{\infty} \left| n(s)-n^{Poisson}(s) \right|^2 ds.
\end{equation}

\begin{figure}[h]
\begin{center}
\leavevmode
\epsfig{file=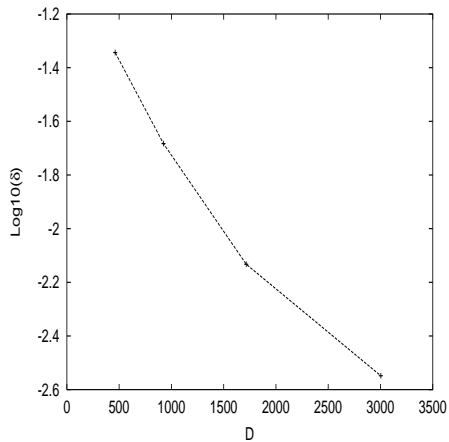,height=6cm,width=6cm,angle=-90}
\caption[]{Logarithmic plot of the average quadratic distance,
$\delta^2$, between the accumulated nearest spacing distribution
obtained from the 200 {\bf TBPRE} members and the Poisson limit, given
by $n^{Poisson}(s) = 1 - \exp{(-s)}$.}
\label{fig:desv6}
\end{center}
\end{figure}

We consider four $TBPRE$ with $(\Omega,N) = (11,6)$, $(12,6)$,
$(13,6)$ and $(14,6)$. The matrix dimensions for these ensembles are
$D = 462$, $924$, $1716$ and $3003$ respectively. In order to have
approximately $2 \times 10^5$ spacings in each of the four ensembles,
different number of members were chosen for each one of
them. Fig. \ref{fig:desv6} shows the logarithm of the average distance
$\delta$ as a function of the matrix dimensionality $D$.  The most
relevant observed feature is that $\delta$ decreases by an order of
magnitude as the dimensionality increases from $D=462$ to
$D=3003$. Moreover, the smooth and monotonous decrease of this
function suggests that it goes to zero in the large $D$ limit.

Summarizing, the use of a new family of fully integrable and exactly
solvable pairing models with a large number of free parameter which
can be selected at random allowed us to perform several stringent
tests of the Berry and Tabor semiclassical proof.  Based on the
numerical results obtained, we conclude that quantum integrable
systems indeed follow a Poisson distribution of nearest neighbor level
spacings for large enough systems.  Exceptions to this rule can be
found, but we showed that they are isolated solutions of high rank
Hamiltonians and that they quickly decay to a Poisson distribution
with an infinitesimal integrable perturbation.

We thank O. Bohigas, P. Leboeuf and G. Sierra for useful
discussions. This work has been supported by grants
BFM2003-05316-C02-02 and BFM2000-0600

%
%

\end{document}